\documentstyle[prd,aps,preprint]{revtex}

\def\spose#1{\hbox to 0pt{#1\hss}}
\def\lta{\mathrel{\spose{\lower 3pt\hbox{$\mathchar"218$}}
     \raise 2.0pt\hbox{$\mathchar"13C$}}}
\def\gta{\mathrel{\spose{\lower 3pt\hbox{$\mathchar"218$}}
    \raise 2.0pt\hbox{$\mathchar"13E$}}}

 \def\Om{{\mit \Omega}}     \def\Ov{{\mit\Omega}_{\rm v}}
 \def\G{{\cal G}}           \def\H{{\cal H}} 
 \def\GEW{\G_{_{\rm EW}}}   \def\GGUT{\G_{_{\rm GUT}}}      
 \def\Gam{{\mit\Gamma}}     \def\GeV{\,{\rm GeV}}

 \def\alp{a}                \def\bet{b}  
             
           \def\zf{{\cal Z}_{\rm f}} \def\z{{\cal Z}}
 \def\ms{m_\sigma}          \def\mx{m_{\rm x}}        \def\mc{m_{\rm c}}
 \def\mP{m_{\rm P}\!}                 
 \def\Ev{E_{\rm v}}         \def\lv{\ell_{\rm v}}     \def\Rv{R_{\rm v}}
 \def\nf{n_{\rm f}}         \def\nv{n_{\rm v}}        
 \def\nua{\nu_\star}        \def\nuf{\nu_{\rm f}}
 \def\rhv{\rho_{\rm v}}     \def\rhb{\rho_{\rm b}}    
 \def\rhc{\rho_{\rm c}}     \def\rhN{\rho_{_{\rm N}}}
 \def\aa{{g^*}}                     
 \def\aas{{g^*_\sigma}}     \def\aaf{{g^*_{\rm f}}}

 \def\q{\zeta}              \def\p{\varepsilon}       \def\ef{\varepsilon}
 \def\LS{{\mit\Sigma}}      \def\LSs{\LS_\sigma}

 \def\Lamb{\Lambda}        
 \def\U{{\cal E}}          \def\T{{\cal T}}          \def\Th{T}   
            
 \def\Tx{\Th_{\rm x}}       \def\Ts{\Th_\sigma}     \def\Tf{\Th_{\rm f}} 
 \def\Tr{\Th_{\rm r}}      \def\Td{\Th_{\rm d}}       
 \def\Ta{\Th_\star}         \def\Tra{\Th_\dagger}   
 \def\TN{\Th_{_{\rm N}}}    \def\TEW{\Th_{_{\rm EW}}}
 \def\TGUT{\Th_{_{\rm GUT}}}

\tighten
\begin{document}

\newcommand{\vp}{\varphi}
\newcommand{\be}{\begin{equation}} \newcommand{\ee}{\end{equation}}
\newcommand{\bea}{\begin{eqnarray}} \newcommand{\eea}{\end{eqnarray}}

\title{Cosmic Vortons and Particle Physics Constraints}

\author{Robert Brandenberger$^{1}$\footnote[1]{rhb@het.brown.edu.}, 
Brandon Carter$^{2}$\footnote[2]{carter@obspm.fr.},
Anne-Christine Davis$^{3}$\footnote[3]{A.C.Davis@damtp.cam.ac.uk.} and 
Mark Trodden$^{4}$\footnote[4]{trodden@ctpa04.mit.edu. ~~
Also, Visiting Scientist, Brown University, Providence, RI. 02912.} }
\smallskip

\address{~\\$^1$Physics Department, Brown University, Providence, RI. 02912, 
USA.}

\address{~\\$^2$D.A.R.C., C.N.R.S, Observatoire de Paris-Meudon, 
92 195 Meudon, France.}

\address{~\\$^3$DAMTP, University of Cambridge, Silver Street, Cambridge, 
CB3 9EW,  UK.}

\address{~\\$^4$Center for Theoretical Physics, Laboratory for Nuclear 
Science and
Department of Physics, \\
Massachusetts Institute of Technology, Cambridge, 
Massachusetts 02139, USA.}

\maketitle

\begin{abstract}

We investigate the cosmological consequences of particle physics theories that
admit stable loops of superconducting cosmic string - {\it vortons}. General
symmetry breaking schemes are considered, in which strings are formed at one
energy scale and subsequently become superconducting in a secondary phase
transition at what may be a considerably lower energy scale.
We estimate the
abundances of the ensuing vortons, and thereby derive constraints on the
relevant particle physics models from cosmological observations. These
constraints significantly restrict the category of admissible Grand Unified
theories, but are quite compatible with recently proposed effects whereby
superconducting strings may have been formed close to the electroweak phase
transition.

\end{abstract}

\vfill

\setcounter{page}{0}
\thispagestyle{empty}

\vfill

\noindent BROWN-HET-1036 \hfill        21 May, 1996.

\noindent DAMTP-96-16

\noindent MIT-CTP-2515 \hfill Submitted to 
{\it Physical Review} {\bf D}

\noindent OBSPM-96014

\noindent hep-ph/yymmdd \hfill Typeset in REV\TeX

\vfill\eject

\baselineskip 24pt plus 2pt minus 2pt

\section{Introduction}
\label{sec:1}

In the past few years it has become clear that topological defects produced in
the early universe may have a considerably richer microstructure than had
previously been imagined\cite{sym rest}. In particular, the core of a defect
acquires additional features at each subsequent symmetry breaking which
preserves the topology of the object. The new microphysics associated with
additional core structure has been exploited by several authors to provide
a new, defect-based scenario for electroweak 
baryogenesis\cite{{BDT},{DMEWBG}}.

The purpose of the present paper is to
constrain general particle physics theories by demanding that the 
microphysics of defects in these models
be consistent with the requirements of the standard cosmology. The basic
idea, due originally to Davis and Shellard\cite{{D&S},{D&S 89},{V&S}}, is as
follows. If a spontaneously broken field theory admits linear topological
defects - {\it cosmic strings} -  which subsequently become superconducting,
then an initially weak current on a closed string loop will automatically tend
to amplify as the loop undergoes dissipative contraction. This current may
become sufficiently strong to modify the dynamics and halt the
contraction so that the loop settles down in an equilibrium state known as a
{\it vorton}.

The population of vorton states produced by such a
mechanism is tightly constrained by empirical cosmological considerations. It
was first pointed out by Davis and Shellard that to avoid obtaining a present
day cosmological closure factor $\Om$ greatly exceeding unity, any theory 
giving rise to stable vorton creation by
superconductivity that sets in during string formation is ruled out if the
symmetry breaking scale is above some critical value.
One of the first attempts to estimate this critical scale\cite{C} indicated 
that it
probably could not exceed that of electroweak symmetry breaking at about $10^2
\GeV$ by more than a few orders of magnitude.  Such strong limits are of
course dependent on the supposition that the vortons are
absolutely stable on timescales as long as the present age of the universe.
However, even if the vortons only survive for a few
minutes, this would be sufficient to significantly affect primordial
nucleosynthesis and hence provide limits of a weaker but nevertheless still
interesting kind\cite{vorton papers}.

In all previous work it was supposed that the relevant superconductivity
sets in during or very soon after the primary phase transition in which the
strings are formed. What is new in the present work is an
examination of the extent to which the limits discussed above are weakened if
it is supposed that superconductivity sets in during a distinct secondary
phase transition occurring at what may be a very much lower temperature than
the string formation scale\cite{DPWP}.

The structure of the paper is as follows. In Section IIA we shall, for
completeness, give a brief introductory review of string superconductivity.
In Section IIB we describe the mechanism of formation of a 
vorton from an originally
distended string loop and in Section IIC we summarize the basic
properties of vorton equilibrium states. In
Section IIIA we first comment on how it can be that the formation scale and
the superconductivity scale can be separated by many orders of magnitude.
We then demonstrate, using a suitably simplified statistical description of 
the string
network, how to estimate the vorton abundance for a generic theory as a 
function of the temperature and the symmetry
breaking scales. In Section IIIB
we apply this procedure to the relatively simple case when the
superconductivity develops during the early period when dissipation is
mainly due to the friction of the ambient medium. In Section IIIC we
go on to treat the more complicated
situation that arises if the superconductivity does not develop until the much
later stage in which dissipation is mainly due to gravitational radiation
and in Section IIID we briefly comment on stability issues.
In Section IV we consider the comparitively weak bounds that are obtained if
it is supposed that the vortons are stable only for a few minutes. Finally, in
Section V we consider the rather stronger bounds that are obtained if the
vortons are of a kind that is sufficiently stable to survive as a constituent
of the dark matter in the universe at the present epoch. We conclude in 
Section VI.

\section{Consequences of Cosmic String Superconductivity}

\subsection{Currents in the Witten model.}

In so far as it has been developed at the present time, the quantitative
theory of vorton structure has been entirely based on the supposition that the
essential features are describable in terms of a simple bosonic
superconductivity model of the kind introduced by Witten\cite{W}. This
category of models consists of spontaneously broken gauged  $U(1) \times
U(1)$ field  theories, which generalise the even simpler category of
spontaneously broken gauged $U(1)$ field theories on which the standard Kibble
description of non superconducting cosmic strings is based. 

The Kibble model
is characterised by a potential $V$ with a quartic dependence on a complex
Higgs field $\phi$ of the familiar form 
${\tilde \lambda}(\vert\phi\vert^2
-v^2)^2$. Here ${\tilde \lambda}$ is a dimensionless coupling 
and $v$ is a mass
scale of order the ``Kibble mass" $\mx$, whose square is identifiable
with the string tension, $\T$, which is (in this model) constant and equal to
the mass per unit length. The string is defined as the region in
which $\vert\phi\vert$ is topologically excluded from its vacuum value as
given by $v\simeq\mx$. 

In addition to $\phi$, the Witten model contains a
second complex scalar field, $\sigma$, and is characterised by a quartic
potential depending on several dimensionless parameters that,  like 
${\tilde \lambda}$,
are assumed to be of order unity. Witten's potential function also 
depends on a second mass
parameter, $\ms$ say, which determines the temperature scale below which
$\sigma$ gives rise to a current carrying condensate on the vortex. In the
Witten model the vortex defects are cosmic strings in which the
tension $\T$ is no longer constant but variable, as a function of the current
magnitude $\vert j\vert$, attaining its maximum (Kibble) value only when the
current magnitude vanishes, so that in general one has $\T\leq \mx^{\,2}$.

In earlier discussions\cite{{D&S},{D&S 89},{V&S},{C},{vorton papers}} of
vorton physics it was implicitly or explicitly  supposed that the magnitude of
$\ms$ was not very different from that of
the original symmetry breaking mass parameter  $\mx$. In that case, the 
formation of
the $\sigma$ condensate could be considered as part of the same symmetry
breaking phase transition as that by which the strings themselves were formed.
Our purpose here is to consider scenarios in which $\ms$ may be very
much smaller than $\mx$, as will occur when successive phase transitions at
two entirely distinct cosmological epochs are involved\cite{DPWP}.  It is only
after the second phase transition that a condensate with amplitude
$\vert\sigma\vert$ and angular variable phase, $\theta$ say, will form on the
string world sheet. There then exists an identically conserved worldsheet
phase current with components

\be 
\tilde j^\alp =
{1\over2\pi}\varepsilon^{\alp\bet}\partial_\bet\theta \ , 
\label{old 1} 
\ee 
where $\varepsilon^{\alp\bet}$ are the components of the antisymmetric unit
surface element tensor induced on the 2-dimensional worldsheet. As well as
this topologically conserved phase current, there will also be a dynamically
conserved particle number current with components given\cite{V&S}
by 

\be 
j_\alp = 2{\tilde\Sigma}(\partial_\alp\theta - e A_\alp) \ , 
\label{old 2}
\ee 
where ${\tilde \Sigma}$ is the surface integral of $\vert
\sigma\vert^2$ over the vortex core cross section. Here $A_\alp$ are the
induced components of the electomagnetic background potential and $e$ is
the coupling constant associated with the carrier field if it
is gauged. When $e$ is non zero $\tilde j^\alp$ will be gauge dependent, 
but $j^\alp$ is physically well defined and will determine a 
corresponding electric surface current given by

\be
I^\alp=ej^\alp\ .
\label{plus 1}
\ee

\subsection{Formation of Vorton States.}

In the case of a closed string loop the conserved surface currents
characterised above will determine a corresponding pair of integral quantum
numbers that are expressible in terms of circuit integration round the loop.
These are given by

\be
N = \oint\tilde j^\alp d\ell_\alp\ ,\hskip 1 cm 
Z = \oint j^\alp d\ell_\alp \ .
\label{old 3}
\ee
where $d\ell_\alp$ are the components of the length element normal to the
circuit in the worldsheet. Note that even when $\tilde j^\alp$ is gauge
dependent, $N$ is well defined. A non conducting Kibble type string loop must
ultimately decay by radiative and frictional drag processes until it
disappears completely. However, since a Witten type conducting
string loop is characterised by the classically conserved quantum 
numbers  $N$ and $Z$, such a loop may be saved from
disappearance by reaching a state in which the energy attains a minimum for
given non zero values of these numbers. 

In view of a widespread
misunderstanding about this point, it is to be emphasised that the existence
of such energy minimising ``vorton" states does not require that the carrier
field be gauge coupled. If there is indeed a non vanishing charge coupling
then the loop will of course be characterised by a corresponding total
electric charge 

\be
Q=\oint I^\alp d\ell_\alp   \ ,
\label{plus 2}
\ee
in terms of which the particle number will be expressible directly as $Z=Q/e$.
However, the important point is
that even in the uncoupled case, for which
$I^\alp$, and hence also $Q$, vanish, the
quantum number $Z$ will nevertheless remain perfectly well defined.

Although the essential physical properties of a vorton state will 
be fully determined
by the specification of the relevant pair of integers $N$ and $Z$, it is not
true that any arbitrary choice of these two numbers will characterise a viable
vorton configuration. This is because the requirement that the strictly 
classical string
description should remain valid turns out to be rather restrictive. To start
with, it is evident that to avoid decaying completely like a non conducting
loop, a conducting loop must have a non zero
value for at least one of the numbers $N$ and $Z$. In fact, one would expect 
that                                                      
both these numbers should be reasonably large compared with unity to diminish
the likelihood of quantum decay by barrier tunneling. 
However, even for moderately large
values of $N$ and $Z$ there will be further restrictions on the admissible
values of their ratio $Z/N$ due to the necessity of avoiding spontaneous 
particle emission as a result of current saturation.

The existence of a maximum amplitude for the string current was originally
predicted by Witten himself\cite{W}.
However, quantitative knowledge about the current
saturation phenomenon remained undeveloped until the appearance of an
important pioneering investigation by Babul, Piran, and Spergel\cite{BPS} who
undertook a detailed numerical analysis of the mechanism whereby the presence
of a current tends to diminish the string tension $\T$. In a non conducting
string, the tension $\T$ is identifiable with the energy per unit length $\U$,
but in the conducting case the diminution of $\T$ is accompanied by an
augmentation of $\U$ such that 

\be
\T\leq \mx^{\,2}\leq \U\ .
\label{plus 3}
\ee
The analysis of~\cite{BPS} provided an empirical ``equation
of state" specifying $\T$ as a non-linear function of the current magnitude
$\vert j\vert$ and hence of the energy density $\U$. The minimum of $\T$, and
hence the maximum allowed value for $\U$, is obtained when the current
amplitude $\vert j\vert$ reaches a critical saturation value with order of
magnitude

\be
\vert j \vert^2\approx \U-\T \lta \ms^{\,2}\ .
\label{plus 4}
\ee
Such knowledge of the equation of state is precisely what is required for 
investigating string dynamics. 

We now apply this to vorton equilibrium
states. The phase angle, $\theta$, is expressible in terms of the
background time coordinate $t$ and a coordinate $\ell$ representing arc length
round the loop by

\be 
\theta = \omega t-k\ell \ , 
\label{old 4}
\ee 
where $\omega$ and $k$ are constants. If the total circumference of the vorton
configuration is denoted by $\lv$ then it can be seen that these constants
will be given in terms of the corresponding quantum numbers by 

\be
\omega=\frac{Z}{2{\tilde \Sigma}\lv}\ ,\hskip 1 cm k=\frac{2\pi N}{\lv}\ .
\label{plus 5}
\ee 
The general condition\cite{C ring} for purely
centrifugally supported equilibrium is that the rotation velocity
should be given by a formula of the same simple form as the one\cite{C 89} for
the speed of extrinsic wiggle propagation, namely 
\be
v^2=\frac{\T}{\U} \ .
\label{plus 6}
\ee 
Note that this relationship applies when the electrical coupling
is absent or, as will commonly be the case, negligible because of
the smallness of the fine structure constant\cite{P ring}.

\subsection{Properties of Vortons.}

The random distribution of initial values of the quantum numbers $N$ and $Z$
leads to the formation of a range of qualitatively different kinds of vorton
states. We shall now briefly review the reasons why it is to be expected that
the most numerous initially will be of approximately {\it chiral} type,
meaning that their rotation velocity is comparable with the speed of light as
given by $v=1$. Note however, that it may be that vortons of the very
different {\it subsonic} type, with $v\ll 1$, will be favoured by natural
selection in the long run.

In the case of a {\it spacelike} current, with $\omega^2<k^2$
(the only possibility envisaged in~\cite{BPS}), the
velocity given by~(\ref{plus 6}) is to be interpreted simply as the
{\it phase speed}: 

\be
v={\omega\over k}={\pi Z\over{\tilde \Sigma} N }\ .
\label{plus 7}\ee
Since we are assuming that $\ms$ is small compared with $\mx$, the saturation
limit (\ref{plus 4}) implies 

\be
\frac{\U-\T}{\T}\ll 1 \ ,
\label{plus 8}\ee 
which obviously,  by (\ref{plus 6}), implies the property of approximate
chirality, $v\simeq1$. It is expected on dimensional grounds that, 
although the
sectional integral ${\tilde \Sigma}$ is a function of the current, it 
will not get
extremely far from the order unity. This behaviour has been
observed numerically in particular cases\cite{BPS}\cite{P} 
(see also \cite{{P},{CP}}). It can
therefore be deduced that approximate chirality requires the
vorton to be characterised by a pair of quantum numbers having roughly
comparable orders of magnitude, 

\be
\vert Z\vert \approx N \ ,
\label{plus 9}
\ee
where we use an orientation convention such that $N$ is always positive. 

On purely statistical grounds one would expect that the two
quantum numbers would most commonly be formed with comparable order of
magnitude, so as to satisfy (\ref{plus 9}), though not necessarily with an
exact ratio small enough to give a spacelike vorton current.
In fact, although the limit (\ref{plus
4}) definitely excludes the possibility of vorton states with $N\gg \vert
Z\vert$, it turns out that there is nothing to prevent the existence of
non-chiral vortons having a current that is not just marginally timelike
($\omega^2>k^2$), but which have $\vert Z\vert\gg N$. In the {\it timelike} 
case
the velocity $v$ in (\ref{plus 6}) is not the phase speed but the {\it current
velocity} given by 

\be
v=k/\omega \ . 
\label{current velocity}
\ee

A preliminary application\cite{Moriand 95} of the
principles described in the following sections suggests that subsonic 
vortons, which will necessarily be characterised by $\vert
Z\vert \gg N$, will initially be formed in much smaller numbers than the
more familiar chiral variety described by (\ref{plus 9}). This means
that if ordinary chiral vortons are sufficiently stable to
survive over cosmologically significant timescales (a question that must be
left open for future research) then they can provide us with much stronger
constraints on admissible particle theories than the more exotic subsonic
variety. In order to keep our discussion as clear and simple as possible, we
shall therefore say no more about subsonic vortons and restrict our
attention to the chiral
variety.

Chiral vortons are only marginally affected by the electromagnetic 
coupling, $e$\cite{P ring}.
Therefore, they can be described by the simplest kind of elastic string 
formalism in which
electromagnetic effects are ignored altogether. This means\cite{C ring} that
the mass energy $\Ev$ of such a vorton will be given in terms
of its circumference $\lv$ by 

\be 
\Ev= \lv(\U+\T)\simeq \lv \mx^{\,2}\ \ . 
\label{old 5}
\ee
In order to evaluate this quantity all that remains is to work out $\lv$. 
In the earliest work on
vorton states it was always presumed that they would be circular, with radius
therefore given by $\Rv=\lv/2\pi$ and angular momentum
quantum number $J$ given\cite{C ring} by $J=NZ$ or equivalently by $J^2=
\U\T\lv^{\,4}/4\pi^2$. Thus, eliminating $J$, one obtains the required
result as

\be 
\lv=(2\pi)^{1/2}\vert NZ\vert^{1/2}(\U\T)^{-1/4}
\simeq(2\pi)^{1/2}\vert NZ\vert^{1/2}\mx^{-1} \ . 
\label{old 6}
\ee 
More recent work\cite{Cam} has established that even in cases
where the vorton configuration is
strongly distorted, the expression~(\ref{old 6})
will remain perfectly valid. Combining this with~(\ref{old 5}), 
and recalling that for chiral vortons $|Z|\approx N$,
we thus obtain a final estimate of the vorton mass
energy as
 
\be
\Ev\simeq(2\pi)^{1/2}\vert NZ\vert^{1/2}\mx\approx N\mx\ .
\label{plus 10}
\ee

The preceding formulae are based on a classical description of the string
dynamics. This is valid only if the length $\lv$ is 
large
compared with the relevant quantum wavelengths, of which the longest is the
Compton wavelength associated with the carrier mass $\ms$. It can be seen from
(\ref{old 6}) that this condition, namely

\be
\lv\gg \ms^{-1} \ ,
\label{minimum}
\ee
will only be satisfied if the product of the quantum numbers $N$ and $Z$ is
sufficiently large. A loop that does not satisfy this requirement will never
stabilise as a vorton. After its length has been reduced to the order of
magnitude (\ref{minimum}) by a classical contraction process, it will
presumably undergo a rapid quantum decay whereby it will finally
disappear completely just as if there  were no current.

\section{The Vorton Abundance}

\subsection{Basic postulates: a scheme based on two mass scales.}

The present analysis will be carried out within the framework of the usual 
FRW model in which the universe evolves in approximate thermal
equilibrium with a cosmological background temperature $\Th$.
The effective number of massless degrees of freedom at temperature $\Th$ is
denoted by $g^*$.
Note that $\aa\approx 1$ at low temperatures but that in the range where
vorton production is likely to occur, from the electroweak scale through to
grand unification, $\aa \simeq 10^2$ is a reasonable estimate.

Any vorton formation processes must occur during the radiation
dominated era which ended when the temperature of the universe dropped 
below $10^{-2} \GeV$ and became effectively transparent. The relevant 
cosmological quantities are the age of the universe, given by 
$t \approx H^{-1}$
where $H$ is the Hubble parameter, and the radiation dominated 
time-temperature
relationship

\be
t\approx {\mP\over\sqrt\aa\Th^2} \ ,
\label{plus 18}
\ee  
where $\mP$ is the Planck mass.

During this cosmological evolution, the particle physics gauge group is
assumed to undergo a series of
successive spontaneous symmetry  breaking phase transitions, expressible
schematically as

\be
\GGUT  \mapsto  \cdots \H \cdots  \mapsto    
 \GEW \mapsto  SU(3)\times U(1) \ . 
\label{old 7}
\ee
Here $\GGUT$ is the ``grand unified" group, $\H$ is some hypothetical
intermediate symmetry group (such as that of the axion phase), and $\GEW$ is
the standard model group $SU(3)\times SU(2)\times U(1)$ or
one of its non-standard (e.g. supersymmetric) extensions.
The role of the Witten model is to
provide an approximate description
of an evolution process dominated by two distinct steps in this chain.  

When a semi-simple symmetry group, $\G$ say, is broken down to a 
subgroup, $\H$ say, the topological criterion for cosmic string formation is 
that the first homotopy group of the quotient should 
be non trivial:

\be
\pi_1\{\G/\H\} \neq 1 \ .
\label{old 8}
\ee
Our primary supposition is that such a process occurs at some particular
cosmological temperature, $\Tx$, which we assume to be of the same order
of magnitude as the relevant Kibble mass scale $\mx$.
This mass scale is interpretable as being of the order of the mass of the
Higgs particle responsible for the symmetry breaking according to the simple
model discussed in the previous section. 

Our next basic postulate is that a current carrying field, characterised by 
the independent mass scale $\ms$,
condenses on the ensuing string defect at a subsequent stage, when the
background temperature has dropped to a lower value, $\Ts$, which we
assume to have the same order of magnitude as the mass scale $\ms$.

The formation of a condensate with finite amplitude characterised by the
dimensionless sectional integral ${\tilde \Sigma}$ does not in itself 
imply a non zero
expectation value for the corresponding local current vector, $j$. However,
one expects that thermal fluctuations will give rise to a non zero value for
its squared magnitude $\vert j\vert^2$ and hence that a random walk process
will result in a spectrum of finite values for the corresponding
string loop quantum numbers $N$ and $Z$. Therefore, in the long run,
those loops for which these numbers satisfy the minimum length
condition (\ref{minimum}) are predestined to become stationary
vortons,
provided of course that the quantum numbers are strictly conserved during the
subsequent motion, a requirement whose validity depends on the condition that
string crossing processes later on are statistically negligible. We describe
these loops as {\it protovortons}.

Note that the protovortons will not become
vortons in the strict sense until a lower temperature, the
vorton relaxation temperature $\Tr$ say (whose value will not be relevant for
our present purpose) since the loops must first lose their excess energy.
Whereas frictional drag and electromagnetic radiation losses will commonly
ensure rapid relaxation, there may be cases in which the only losses
are due to the much weaker mechanism of  gravitational radiation. 

As the string network evolves, the distribution rarifies due to damping out 
of its fine structure
first by friction and later by radiation reaction. However,  not all of its 
lost energy
goes directly into the corresponding frictional heating of the background or
emitted radiation. There will always be a certain fraction, $\ef$ say, that 
goes into loops which evolve without subsequent
collisions with the main string distribution.
It is this process that provides the raw material for vorton production.
Such loops will ultimately be able to survive as 
vortons if the current induced by random fluctuations during the carrier
condensation process  is sufficient for the condition    (\ref{old 6}) to be
satisfied, i.e. provided its winding number and particle number are large
enough to satisfy

\be 
\vert NZ \vert^{1/2} \gg {\Tx\over\Ts}\ .
\label{minprod}
\ee
Any loop that fails to satisfy this
condition is doomed to lose all its energy and disappear.

In favorable circumstances, (namely those considered in Section IIIB) most of
the loops that emerge in this way at $\Ts$ will satisfy the 
condition~(\ref{minprod}) and thus be describable as
protovortons. However in other cases (namely those considered in Section IIIC)
the majority of the loops that emerge during the period immediately following
the carrier condensation will be too small to have aquired sufficiently large
quantum numbers by this stochastic mechanism. These loops will therefore 
not be viable
in the long run and are classified as {\it doomed loops}. 
Nevertheless, even in such
unfavourable circumstances, the monotonic increase of the damping length
$L_{\rm min}$ will ensure that at a lower temperature $\Tf<\Ts$ a later, and 
less prolific, generation of emerging loops will after all
be able to qualify as protovortons. We refer to $\Tf$ as the protovorton
formation temperature.

The scenario summarised above is based on the accepted understanding of the
Kibble mechanism\cite{V&S}, according to which, after the temperature has
dropped below $\Tx$ the effect of various damping mechanisms will remove most
of the structure below an effective smoothing length, $L_{\rm min}$, which will
increase monotonically as a function of time, so that nearly all the surviving
loops will be have a length $L=\oint d\ell$ that satisfies the inequality 

\be
L\gta L_{\rm min}
\label{loopl}
\ee
There will thus be a distribution of string loops, of which the most numerous
will be relatively short ones, with $L\approx L_{\rm min}$, that are on the 
verge 
of
emerging, or that have already emerged, as protovortons or doomed loops
as the case may be.

Whereas on larger scales closed loops and wiggles on very long string segments
will be tangled together, on the shortest scales, characterised by the lower
cutoff $L_{\rm min}$, loops will be of a relatively smooth form.
It is these smallest loops that are
candidates for subsequent transformation into vortons.

The total number density of small loops with length and radial
extension of the order of $L_{\rm min}$ will (due to the rapid fall off of the 
spectrum
that is expected for larger scales) be not much less than the number density
of all closed loops and so will be given by an expression of the
form

\be
n\approx \nu\  L_{\rm min}^{-3}
\label{plus 22}
\ee
where $\nu$ is a time-dependent parameter which we will discuss later.

The theory reviewed above was originally developed on the assumption 
that the string evolution is governed by  Goto-Nambu
type dynamics. In the kind of scenario we are considering, this condition will
obviously be satisfied as long as the cosmological temperature $\Th$ is
greater than or comparable with the carrier condensation temperature $\Ts$.
Moreover, the usual Goto-Nambu type description, and its consequences as
described above, will remain valid for a while after the strings have
become superconducting since the
currents will initially be too weak to have significant dynamical effects.
The Goto-Nambu theory inevitably breaks down at some temperature above $\Tr$.
However, there may be cases for which such a description will break down 
even before the protovorton formation temperature $\Tf$ is reached.

The typical length scale of
string loops at the transition temperature, 
$L_{\rm min}(\Ts)$, is considerably greater than relevant thermal 
correlation
length, $\Ts^{-1}$, that will presumably characterise the local current
fluctuations at that time. It is because of this that string loop evolution is
modified after current carrier condensation.
The inequality

\be
L_{\rm min}(\Ts)\gg \Ts^{-1}  \ ,
\label{plus 24}
\ee
and the fact that, by (\ref{loopl}), the length of any loop present at the 
time
of the condensation will satisfy $L\gta L_{\rm min}(\Ts)$, means that the 
random walk
effect can build up reasonably large, and typically comparable initial values
of the quantum numbers $\vert Z\vert$ and  $N$. The reason is that for a loop
of length $L$, the expected root mean square values produced
in this way from carrier field fluctuations
of wavelength $\lambda$ can be estimated as

\be
\vert Z\vert \approx N \approx \sqrt{L\over\lambda} \ . 
\label{N}
\ee
At the time of the condensation,
a typical loop is characterised by

\be
L\approx L_{\rm min}(\Ts)\ 
\label{typil}
\ee
and

\be
\lambda\approx\Ts^{-1}
\label{lambda}
\ee
so that one obtains the estimate

\be
\vert Z\vert \approx N \approx \sqrt{L_{\rm min}(\Ts)\Ts} \ , 
\label{plus 29}
\ee
which, by~(\ref{plus 24}), is large compared with unity.

For current condensation during the friction dominated regime discussed in
Section IIIB, we shall see that this will always be sufficient to satisfy
the requirement (\ref{minprod}). However, this condition will not hold for
condensation later on in the radiation damping regime discussed in Section
IIIC. In the latter case, typical small loops that free themselves from the
main string distribution at or soon after the time of current condensation
will be doomed loops
since they do not satisfy~(\ref{minprod}).
However, there will always be a minority of longer
loops for which~(\ref{minprod}) will be satisfied, namely those exceeding a 
minimum length given, according to (\ref{plus 29}), by

\be
L\approx {\Tx{\,^2}\over \Ts^{\,3}} \ .
\label{longenuf}
\ee   
This condition is still not quite
sufficient to qualify them as protovortons since such exceptionally long loops
will be very wiggly and collision prone. It is not until a
later time at a lower temperature $\Tf$ that free protovorton loops will
emerge. In this case, the
typical wavelength of the carrier field will be given by~(\ref{lambda}) and
the final value of the length of a typical loop is

\be L\approx L_{\rm min}(\Tf) \ .
\label{xif}
\ee
The new estimate for the values of the quantum numbers is

\be
\vert Z\vert \approx N \approx \sqrt{L_{\rm min}(\Tf)\Ts\over \zf} \ , 
\label{Nlater}
\ee
where we have included a blueshift factor, $\zf$, whose value is not 
immediately obvious
but that is needed allow for the net effect on the string of weak stretching
due to the cosmological expansion and stronger shrinking due to wiggle damping
during the period as the temperature cools from $\Ts$ to $\Tf$. In
the earlier friction dominated regime $\Tf$ is identifiable with $\Ts$ so the
problem does not arise, and in the radiation damping era
cosmological stretching  will in fact be
negligible. Therefore, the net effect is that $\zf$ will be small compared
with unity, the hard part of the problem being to estimate how much so.

The value given by (\ref{Nlater}) will
increase monotonically as $\Tf$ diminishes. 
The required value of $\Tf$, at which the
formation of the protovorton loops will actually occur, is that for which
the function in~(\ref{Nlater}) reaches the minimum qualifying value
given by~(\ref{minprod}). This value is thus obtainable in principle by
solving the equation

\be
{L_{\rm min}(\Tf)\over \zf} \approx {\Tx^2\over \Ts^3}\ ,
\label{fff}
\ee
but this can only be done in practice when we have found the $\Tf$-dependence
of $L_{\rm min}(\Tf)$ and $\zf$. We discuss the $\Tf$-dependence of $\zf$ 
shortly. 

The number density of protovorton loops at the
temperature $\Tf$ will be comparable with the total loop number density at the
time, so that by (\ref{plus 22}) it will be expressible as

\be
\nf\approx\ef\nuf\, L_{\rm min}(\Tf)^{-3} \ ,
\label{newf}
\ee
where $\ef$ is an efficiency factor of order unity, and $\nuf$ is the
value of the dimensionless parameter $\nu$ at that time. If the current
condenses in the friction dominated regime
$\nuf$ will
simply have an order of unity value.
However, if the condensation  does not occur until later on, in the
radiation dominated era,  $\nuf$ will have a lower value which is not so
easy to evaluate.

The number of protovorton loops in a comoving
volume will be approximately conserved during their subsequent 
evolution. Therefore, since volumes will
scale proportionally to the inverse of the entropy density
it follows that the number density $\nv$ of the resulting vortons at
a lower temperature $\Th$ will be given in terms of the number density $\nf$
of the proto-vorton loops at the time of condensation by

\be
{\nv\over\nf}\approx f\left( {\Th\over\Tf}\right)^{3} \ .
\label{add 1}
\ee
Here $f$ is a dimensionless adjustment factor that we expect to be small but
not very small compared with unity, and that will be given by

\be
f\simeq{\ef\aa\over\aaf} \ ,
\label{plus 27}
\ee
where $\aaf$ is the value of $\aa$ at the protovorton formation temperature
$\Tf$.
          
Using~(\ref{plus 10}) the corresponding mass density
will be given by 

\be
\rhv\approx N\mx \nv\ . 
\label{plus 28}
\ee 
Thus, the mass density
of the distribution of the protovortons in the range $\Tf \gta \Th \gta \Tr$,
and of the mature vortons after their formation in the range $\Th \lta \Tr$, 
is given by the general formula

\be
\rhv \approx f\nuf {\Tx\Ts^{1/2}\over\zf^{1/2}L_{\rm min}(\Tf)^{5/2}}
\Big({T\over\Tf}\Big)^3 \ .
\label{add 2}
\ee
 
When the dependences of $\nuf$, $L_{\rm min}(\Tf)$ and $\Tf$ on the fundamental 
parameters
$\Tx$ and $\Ts$ are known, the formula (\ref{add 2}) will allow us to place
limits on $\Ts$ by determining how the presence of the corresponding
population of remnant vortons would affect the course of cosmic evolution. In
the following sections we shall derive several constraints on such a
population by demanding that it not significantly interfere with the
cornerstones of the standard cosmology. However, before we can do so it
remains to obtain at least rough estimates of the required values of the
dependent variables. This turns out to be fairly easy in the case of
condensation during the friction dominated era that will be discussed in the
next section. However the derivation of firm conclusions is less
straightforward for the kind of scenario discussed in Section IIIC, in which
the current condensation occurs in the radiation dominated regime.

\subsection{Condensation in the friction damping regime.}

According to the standard picture\cite{KEH},
the evolution of a cosmic string network
is initially dominated by the frictional drag of the thermal background.
The relevant dynamical damping timescale, $\tau$,
during this period is approximately given by 

\be
\tau\approx{\Tx^{\,2}\over\beta\,\Th^3} \ , 
\label{plus 19}
\ee
where $\beta$ is a dimensionless drag coefficient that depends on the details
of the underlying field theory but that is typically
expected\cite{{KEH},{V&V}} to be of order unity. In this regime 
the large scale structure is frozen and
retains the Brownian random walk form~(\ref{plus 22}). However,
the microstructure is smoothed out below a correlation length $L_{\rm min}$
given by
 
\be 
L_{\rm min}\approx\sqrt{\tau t} \ ,
\label{plus 20}
\ee
where $t$ is the Hubble time.
Neglecting the very weak $g^*$-dependence, the
required correlation length is thus found to be given by 

\be
L_{\rm min}\approx 
\left({\mP\over\beta}\right)^{1/2}{\Tx\over\Th^{\,5/2}}\ . 
\label{old 13}
\ee

The friction-dominated regime continues until the temperature $\Th$ 
drops below a critical value $\Ta$ given by 

\be
\Ta \approx\frac{\Tx^{\,2}}{\beta\,\mP} \ ,
\label{old 10}
\ee
at which $\tau$ is comparable with $t$.

Setting $T$ equal to $\Ts$ in (\ref{old 13}) in order to obtain the relevant
value of $L_{\rm min}(\Ts)$, and using (\ref{plus 29}) and  (\ref{typil}), the 
required
expectation value for the quantum number $N$ can be
estimated as

\be
N\approx\left({\mP\over\beta\Ts}\right)^{1/4}
\left({\Tx\over \Ts}\right)^{1/2} \ .
\label{old 21}
\ee
It follows from  (\ref{plus 10}) and (\ref{old 6}) that a typical vorton in
this relic distribution will have a mass-energy given by 

\be
\Ev\approx\left({\mP\over\beta\Ts}\right)^{1/4}
\left({\Tx^3\over\Ts}\right)^{1/2} \ ,
\label{plus 30}
\ee                    
which corresponds to a vorton circumference 

\be
\lv\approx \left(\frac{\mP}{\beta\Ts}\right)^{1/4}\big(\Tx\Ts\big)^{-1/2} \ .
\label{old 20}
\ee
It can thus be confirmed using (\ref{old 10}) that the
postulate  $\Ts > \Ta$ automatically ensures that these vortons will indeed
satisfy the minimum length requirement (\ref{minimum}), though only
marginally when $\Th$ is at the lower end of this range.

From (\ref{newf}) and (\ref{old 13}) the number density of these
proto-vorton loops, at formation, is 

\be
\nf\approx \nua\left({\beta\,\Ts\over\mP}\right)^{3/2}
\left({\Ts^{\,2}\over\Tx}\right)^3\ .
\label{plus 25}
\ee
It follows from (\ref{add 1}) that at 
later times the number density of their mature vorton successors will be 

\be
\nv\approx\nua f\left({\beta\Ts\over \mP}\right)^{3/2}
\left({\Ts\Th\over\Tx}\right)^3 \ . 
\label{plus 26}
\ee
Thus, after the temperature has fallen below the value $\Tr$, the resulting 
mass density of the relic vorton population will be

\be
\rhv\approx \nua f N\left({\beta\Ts\over\mP}\right)^{3/2}\,
\left({\Ts\over\Tx}\right)^2 \Ts\Th^3\ , 
\label{old 17}
\ee 
which by (\ref{old 21}) gives our final estimate as

\be
\rhv\approx \nua f\left({\beta\Ts\over\mP}\right)^{5/4}
\left({\Ts\over\Tx}\right)^{3/2}\Ts\Th^3 \ . 
\label{plus 31}
\ee

\subsection{Condensation in the radiation damping regime.}

For strings formed at low energies, for example in some non-standard
electroweak symmetry breaking transition, the scenario of the
preceding subsection is the only one that needs to be considered. However, for
strings formed at much higher energies, in particular for the commonly
considered  case of GUT strings, current condensation could occur
during the extensive temperature range
below $\Ta$. The
minimum length requirement is only marginally satisfied by typical loops when
condensation occurs near the end of the friction dominated regime. Therefore,
if $\Ts < \Ta$, typical loops present during the 
transition will not
be long enough to qualify as protovortons. This means that the vorton
formation temperature $\Tf$ will not coincide with $\Ts$ as it did in the
friction dominated regime, but rather will have a distinctly lower value.

In these scenarios the final stage of protovorton formation will be preceeded
by a period of evolution in the temperature range $\Ta\gta \Th\gta\Tf$.
During this interval friction will be negligible and the only significant 
dissipation
mechanism will be that of radiation reaction. Moreover, during the first part
of this period, in the range $\Ta > \Th >\Ts$, the only radiation mechanism
will be gravitational, which is so weak that to begin with it will have no
perceptible effect at all. Thus, there will be an interval during which the
smoothing length remains roughly constant at its value 
at the end of the friction dominated era, given, according to
(\ref{old 13}) and (\ref{old 10}), by

\be
L_{\rm min}(\Ta)\approx {\beta^2\mP^3\over\Tx^4} \ .
\label{xicrit}
\ee

During the last stage before the protovortons are formed, in the range
$\Ts>\Th>\Tf$ there will already be currents on the strings.
However, in practice, even in
the coupled case the expected currents will be too weak for electromagnetic
radiation damping to be
important. Therefore, gravitational radiation is the only important effect 
throughout the range $\Ta>\Th>\Tf$.

The resulting gravitational smoothing scale will be the length of the
shortest loop for which the survival time exceeds the cosmological timescale
(\ref{plus 18}). From dimensional considerations this can be estimated by 
the expression

\be
t\approx {L_{\rm min}\over\Gam G\,\U}\ ,
\label{add 5} 
\ee
where $\U\simeq \T \simeq \Tx^{\,2}$ is the mass energy density of the string.
Here $\Gam$ is a dimensionless coefficient of order unity and, for
the GUT strings the
gravitational factor will be given by $G\,\U\simeq (\mx/\mP)^2\approx
10^{-6}$. The validity of the formula (\ref{add 5}) has been confirmed in many
particular cases by numerical simulations\cite{V&S}, though the value of the
coefficient turns out to be typically $\Gam\approx 10^2$.
 
Equating (\ref{add 5}) to the cosmological timescale we have

\be
L_{\rm min}\approx{\Gam\over\sqrt\aa\,\mP}\left({\Tx\over \Th}\right)^2 \ .
\label{add 7}
\ee 
This formula is valid when its value becomes larger than that given
by~(\ref{xicrit}). This occurs at a critical value $\Tra<\Ta$
which, from (\ref{add 7}), is given by

\be
\Tra\approx \Big({\Gam\over\sqrt\aa}\Big)^{1/2} 
{\Tx^{\,3}\over\beta\,\mP^{\,2}} \ .
\label{Trad}
\ee

The relation (\ref{add 7}) may also be expressed as

\be
L_{\rm min}\approx\kappa t
\label{defalpha}
\ee
where $\kappa$ is a constant given by

\be
\kappa\approx \Gam \Big({\Tx\over\mP}\Big)^2 \ .
\label{alpha}
\ee
To avoid ambiguity we can of course simply use the formula
(\ref{defalpha}) as a defining relation to specify the parameter $\kappa$
during the Hubble damping ``doldrum" regime $\Ta\gta\Th\gta\Tra$.
However, with this convention $\kappa$ will have to be considered as a 
function of
time, starting with unit value at $\Ta$ and decreasing to the very low 
value (\ref{alpha}) at which it levels off at $T\approx \Tra$.
This description of the network evolution is illustrated graphically
in figure~1.

In order to apply the formula
(\ref{add 2}) for the final vorton mass density we must evaluate
the dimensionless
coefficient $\nu$ that determines the protovorton number density. It was easy
to do this for the friction dominated case for which we could assume
a constant value $\nua$ of the order of unity. However, $\nu$ may
subsequently be reduced by an amount whose estimation is not so clearly
evident. It is reasonable to expect that in the radiation dominated
regime the string distribution tends towards a scale invariant form, albeit
one that will not be quite so simple as the
Brownian form that prevailed in the friction-dominated era.
Although there may already be approximate scaling
for large scales, it is clear that scaling on the smaller scales that matter
for our present purpose can only be obtained after the parameter $\kappa$
defined by (\ref{defalpha}) has settled down to a constant value. This
occurs at
temperatures below $\Tra$ for which a detailed analysis is beyond the scope of
the simulations that have been achieved so far. For simplicity we shall
use a crude, but we hope sufficiently robust, description based on simple
and quite natural physical considerations.

In so far as $\nu$ is concerned, the plausible
conjecture that it should be describable by a scaling solution is to be
interpreted as meaning that it should be a function only of the dimensionless
ratio $R/t$ (where, it is to be recalled, $R$ denotes the radial scale
under consideration and $t$ is the Hubble time).
It is reasonable to expect that for values of $R$ in the range
$\kappa\ll R/t \ll 1$, the value of $\nu$ should be given by a 
simple power law of the form

\be
\nu\approx\nua \Big({R\over t}\Big)^\q \ ,
\label{scaling}
\ee
with constant index $\q$.
We expect that, in the radiation dominated regime, the appropriate value of
the index should be close to but perhaps slightly greater than a lower
limit given by\cite{T&B 86}

\be
\q={3\over 2} \ .
\label{index}
\ee
(The analogue for the matter era in which we are situated today is a value
slightly greater than a lower limit given by $\q=2$.) 

Assuming that the formula (\ref{scaling}) still gives the right order of
magnitude at the lower end of its range, $\nu$ will be
given in the radiation dominated regime by the constant
value

\be
\nu \approx \nua\, \kappa^\q
\label{barnu}
\ee
with $\kappa$ given by $(\ref{alpha})$. This means that the corresponding
value of the loop number density itself will be given according to
(\ref{defalpha}) by

\be
n\approx \nua\, \kappa^{\q-3} t^{-3} \ .
\label{nloop}
\ee

Before evaluating the required result (\ref{add 2}), it remains to
obtain the value $\Tf$.
To do this we have to solve the equation (\ref{fff}) that results
from the minimum length requirement, which, from (\ref{add 7}),
reduces to

\be
\zf\Tf^{\,2}\approx{\Gam\Ts^{\,3}\over\sqrt\aa\,\mP} \ .
\label{quation}
\ee 
However, before we can solve this deceptively
simple equation in practice, we need to know the $\Th$ dependence of the
factor $\z$. This is the most delicate part of the calculation,
since it involves competing effects of comparable magnitude.   

As the string distribution evolves, the time dependent blue shift factor
$\z$ is the factor by which the supporting string length has shrunk
since the time of the condensation at the temperature $\Ts$. This shrinking
can be accounted for as the net result of three main effects,
two of which are comparitively easy to evaluate. 

The weakest of these effects is the stretching due to the expansion
of the universe. This will always be more than compensated (except in the
``doldrum" period in which compensation is
only marginal) by shrinking due to the steady
damping out of the short wavelength wiggles that give the most important
contribution to the total string length per unit volume in the
radiation dominated era. If these two effects
were the only ones it would be relatively easy to estimate $\z$ since it would
simply be proportional to the total string length, ${\LS}$ say, in a comoving
volume that can conveniently be taken to be a cubic thermal wavelength.
This is given by

\be
\LS\approx  \Lamb\Th^{-3}
\label{longsum}
\ee
where $\Lambda$ is the total string length per unit volume.
Now, note that the main contribution to the total 
length of the string distribution is provided
by short wavelength modes with scale of order the smoothing length
$L_{\rm min}$. Therefore, $\Lamb$ can be estimated as
being $nL_{\rm min}$, where $n$ is given by
(\ref{plus 22}). This implies

\be
\Lambda\approx \nu\, L_{\rm min}^{-2} \ ,
\label{Lamb}
\ee
in which $\nu$ is given by (\ref{barnu}).

If the only effect of the damping were to smooth out the short wavelength
wiggles on the main part of the string distribution, it would be possible to
identify $\z$ with the  ratio of $\LS$ to its value $\LSs$ at the time of the
current condensation. However, this would not
allow for a third important effect, namely the losses of 
small
loops which are continually liberated from the string distribution
at the lower end of the spectrum.
Whether these loops survive as vortons or disappear altogether, the effect
on the main part of the string distribution
is that the corresponding string lengths must be subtracted at each stage.
Thus, the total shrinking factor $\z$ will be the ratio of $\LS $, not to
its original known value $\LSs$,  but to a value that is
considerably reduced in such a way as to take account of
this new effect.

In terms of the variation $\delta\LS$ of $\LS$, the variation
$\delta\z$ of $\z$ is given by

\be
{\delta\z\over\z} \simeq { \delta\LS + \Delta\LS \over \LS}
\label{looploss}
\ee
where $\Delta\LS$ is the length of string irreversibly chopped off in the form
of small loops per comoving thermal volume within the short time
interval $\delta t$ under consideration.  The delicate question is that of
quantifying $\Delta\LS$. It is not hard to obtain an order of
magnitude estimate, but since the final result is rather sensitively dependent
on this quantity a more accurate estimate would be desirable. 
In the absence of a detailed numerical investigation, we express 
$\Delta\LS$ as a fraction of the total lost length 

\be
\Delta\LS \simeq -\p   \delta\LS \ ,
\label{efficiency}
\ee
where $\p$ is a dimensionless efficiency factor that must lie in the
range $0 <\p < 1$. This factor is roughly identifiable with the
coefficient introduced, using the same notation, in (\ref{newf}). However in
that context it was sufficient to know that it should be of order
unity.

Whatever the exact value of $\p$, the substitution
of (\ref{efficiency}) in (\ref{looploss}) provides a differential equation
that can be solved to give

\be
\z\approx\Big({\LS\over\LSs}\Big)^{1-\p} 
\label{shift}
\ee
and using~(\ref{add 7}) and (\ref{Lamb}) we finally obtain

\be
\z\approx\Big({\Th\over\Ts}\Big)^{1-\p} \ .
\label{blue}
\ee
It is to be remarked that if the efficiency $\p$ of loop production
were zero, this would mean that the carrier field would be blue shifted
by a factor that would be precisely the inverse of that by which the 
background
radiation is redshifted. However, if substantial loop production occurs
there will be a blueshift by a moderate factor.

Assuming the particle number
weighting factor can be taken to have the
fixed value $\aas$,~(\ref{quation}) and~(\ref{blue}) then give

\be
{\Tf\over\Ts}=\Big({\Gam\,\Ts\over\sqrt\aas\,\mP}\Big)^{1/(3-\p)} \ .
\label{solution}
\ee

Using this result in conjunction with the estimates (\ref{add 7}) and
(\ref{barnu}) the formula (\ref{add 2}) gives the mass density of the 
resulting vorton distribution as

\be
\rhv\approx\ef\aa\nua\Gam^{\q-5/2}\Big({\mP\over\Tx}\Big)^{5-2\q}
\Big({\Ts\over\mP}\Big)^{5/2}
\Big({\Gam\,\Ts\over\sqrt\aas\,\mP}\Big)^
{(3+\p)/(6-2\p)}\,
\Tx\Th^3 \ .
\label{fmassden} 
\ee
If we adopt the value (\ref{index}) for $\q$, this 
simplifies to

\be
{\rhv\over\Tx\Th^3}\approx\ef\aa\nua\Gam^{-1/2}\Big({\mP\over\Tx}\Big)^{2}
\Big({\Ts\over\mP}\Big)^{5/2}
\Big({\Gam\,\Ts\over\sqrt\aas\,\mP}\Big)^
{(3+\p)/(6-2\p)}\ ,
\label{den3} 
\ee
but there remains an unsatisfactory degree of sensitivity to the uncertain
efficiency factor $\p$. This is because although this index will always be 
quite small, the factor $\Ts/\mP$ is very tiny in the cases of
interest.

\subsection{Stability Issues}

Before we consider cosmological constraints we would like to say a few words
about stability. One of our postulates is that superconducting  current
conservation is sufficiently effective to allow the protovortons that emerge
at the temperature $\Tf\lta\Ts$ to settle down as dynamically stable bodies, 
at
a possibly lower relaxation temperature $\Tr\lta\Tf$. This does not exclude 
the
possibility that in the very long run they may finally decay by quantum
tunneling or other ``secular" instability mechanisms. In this case they would
ultimately disappear when the thermal background reached a corresponding
vorton death temperature with an even lower value, $\Td$ say. This means that
a complete analysis of vorton formation and evolution could involve five
successive temperature scales related by

\be
\Tx \gta \Ts \gta \Tf \gta\Tr \gta \Td \ .
\label{old 9}
\ee

Protovortons
are small compared with the ever expanding
scales characterising the rest of the string distribution and hence
undergo  few extrinsic collisions. Also, they are sufficiently 
smooth to avoid destructive fragmentation by self collisions. It therefore
follows
that in most cases the relevant quantum numbers $N$ and $Z$ will be 
conserved. As a
consequence, the statistical properties of the future vorton population will
be predetermined by those of the corresponding protovorton loops at the time
of their emergence at the temperature $\Tf$. It is therefore 
unnecessary for our present purpose to consider how long it takes for the
protovorton loops to settle down and become proper stationary vortons.
Thus, the value of $\Tr$ will not play any role in the discussion that
follows. This is convenient because the details of protovorton loop decay
have not yet been adequately studied, and will obviously be
sensitively dependent on whether the current is electromagnetically coupled,
in which case would expect the later stages of the protovorton loop 
contraction to be relatively rapid.

The final decay temperature $\Td$ is also absent from the
quantitative formulae that we will derive. Indeed, its only role is to
characterise the two principle kinds of scenario that we consider.
In Section V, we discuss vortons which
survive until the present epoch, which requires that $\Td$ should not much
exceed $10^{-12} \GeV$ (corresponding to the observed 3 degree radiation
temperature). However, in Section IV we adopt the weaker
supposition that the vortons survive at least till the time of
nucleosynthesis, which requires only that $\Td$ should not much exceed
$10^{-4} \GeV$. This means that the only temperature scales that remain
as variable parameters in the analysis that follows are the Higgs-Kibble
temperature $\Tx$, the condensation temperature $\Ts$, and the protovorton 
formation temperature $\Tf$ which will not be truly independent but is
a function of $\Ts$ and $\Tx$.

\section{The Nucleosynthesis Constraint.}

One of the most robust predictions of the standard cosmological model is the
abundances of the light elements that were fabricated during primordial
nucleosynthesis at a temperature $\TN\approx 10^{-4} \GeV$.

In order to preserve this well established picture, it is necessary that the
energy density in vortons at that time, $\rhv(\TN)$ should have been small
compared with the background energy density in radiation, 
$\rhN\approx\aa\TN^4$. 
Assuming that carrier condensation occurs during the
friction damping regime and that $\aa$ has dropped to a value of order unity
by the time of nucleosynthesis, it can be seen from (\ref{plus 31}) that this
restriction,

\be
\rhv(\TN) \ll \rhN \ ,
\label{old 22}
\ee
is expressible as

\be
\ef\nua\aas^{-1}\beta^{5/4}\mP^{-5/4}\Tx^{-3/2}\Ts^{\,15/4}\ll\TN\ .
\label{old 24}
\ee 
Below we apply this constraint to some specific examples.

\subsection{ Case $\bf \Tx \approx\Ts$.}

The case for which carrier condensation occurs at or very soon after the
time of string formation has been studied previously and yields
rather strong restrictions for very long lived vortons \cite{C}. If
it is only assumed that the vortons survive for a few minutes, which is all
that is needed to reach the nucleosynthesis epoch we obtain a much weaker
restriction.
Setting $\Ts$ equal to $\Tx$ in (\ref{old 24}) gives

\be
\left(\frac{\ef\nua}{\aas}\right)^{4/9}\Tx \ll 
\left(\frac{\mP}{\beta}\right)^{5/9}\TN^{4/9}\ .
\label{old 25}
\ee
Taking $\aas
\approx 10^2$ and assuming (in view of the low value of the index) that the
net efficiency factor $(\ef\nua)^{4/9}$ and the drag factor $\beta^{5/9}$ are
of the order of unity yields the inequality 

\be
\Tx\lta 10^9\ \GeV \ .
\label{plus 34}
\ee
This is the condition that must be satisfied by the formation temperature 
of {\it cosmic strings that become superconducting immediately}, subject to 
the rather
conservative assumption that the resulting vortons last for at least a few
minutes. It is to be observed that this condition rules out the formation of
such strings during any conceivable GUT transition, but is consistent by a
wide margin with their formation at temperatures close to that of the 
electroweak symmetry breaking transition.

\subsection{Case $\bf \Tx\simeq\TGUT\approx10^{16}$ GeV.}

Here we wish to calculate the highest temperature at which GUT strings can 
become
superconducting without violating the nucleosynthesis constraints. Setting
$\Tx$ equal to $\TGUT$ in (\ref{old 24}), and again using $\aas\approx 10^2$,
we obtain

\be 
\Ts \lta (\ef\nua)^{-4/15}\big(\aas\TN\big)^{\,4/15} \TGUT^{\ 2/5}
\left(\frac{\mP}{\beta}\right)^{1/3} \approx  10^{12}\ \GeV \ , 
\label{old 26}
\ee
where, in the last step, we have neglected the dependence on 
order of unity quantities. It can be
checked, using the Kibble formula (\ref{old 10}), that the maximum value given
by (\ref{old 26}) is at least marginally consistent with the assumption that
current condensation occurs in the friction-dominated regime.
The validity of our derivation is thereby confirmed. It follows
that the nucleosynthesis constraint will always be satisfied when $\Ts$ lies
in the radiation damping epoch. 

Therefore, subject again to the rather conservative assumption that the
resulting vortons last for at least a few minutes, theories in which GUT
cosmic strings  become superconducting above $10^{12}$ GeV are inconsistent
with the observational data.

\section{The Dark Matter Constraint.}

In this section we consider the rather stronger constraints that can be 
obtained if at
least a substantial fraction of the vortons are sufficiently stable to last
until the present epoch.
It is generally accepted that the virial equilibrium of galaxies and
particularly of clusters of galaxies requires the existence of a cosmological
distribution of ``dark" matter. This matter must have a density 
considerably in excess of the baryonic matter
density, $\rhb\approx 10^{-31}$ gm/cm$^3$. On the other hand, on the same
basis, it is also generally accepted that to be consistent with the
formation of structures such as galaxies it is necessary
that the total amount of this ``dark" matter should not greatly exceed the
critical closure density, namely

\be
\rhc\approx 10^{-29} {\rm gm \ cm^{-3}} \ .
\label{add 15}
\ee
As a function of temperature, the critical density scales like the entropy 
density so that it is given by

\be
\rhc(T)\approx \aa\mc\Th^3\ ,
\label{plus 35}
\ee
where $\mc$ is a constant mass factor. Since $\aa\approx1$ at the present
epoch, the required value of $\mc$ (which is roughly as the
critical mass per black body photon) can be estimated as

\be
\mc\approx 10^{-28}\mP\approx 1\ \hbox{eV}\  .
\label{plus 36}
\ee
However, for comparison with the density of vortons that were formed as a
result of current condensation at an earlier epoch characterised by
$\Ts$, what one needs is the corresponding factor $\aas\mc$,
which can be estimated to be

\be
\aas \mc\approx 10^{-26}\mP\approx 10^2\,\hbox{eV}\ .
\label{plus 37}
\ee
(This distinction was obscured in the previous
derivations of the dark matter constraint\cite{C}\cite{Moriand 95}, in which
the value quoted for $\mc$ should be interpreted as meaning the value of
$\aas\mc$, which is what one actually needs.)

The general dark matter constraint is

\be
\Ov \equiv {\rhv\over\rhc}\lta 1\ .
\label{plus 38}
\ee

In the case of vortons formed as a result of condensation during the friction
damping regime
the relevant estimate for the vortonic dark matter fraction is obtainable
from (\ref{plus 31}) as

\be
\Ov\approx\beta^{5/4}\left({\ef\nua\mP\over\aas\mc}\right)
\left({\Ts\over\mP}\right)^{9/4} \left({\Ts\over\Tx}\right)^{3/2} . 
\label{old 27}
\ee
In particular, this formula applies to the case in which the carrier
condensation occurs very soon after the strings themselves are formed, as was
supposed in earlier work. 

However, if we want to strengthen the nucleosynthesis limit (\ref{old 26}) for
the general category of strings formed at the GUT scale, then we are
obliged to consider the case of vortons formed as a result of condensation
during the gravitational radiation damping regimes.
In this case, equation~(\ref{den3}) gives the relevant estimate for the 
vortonic
dark matter fraction as 

\be
\Ov\approx\ef\nua\Gam^{-1/2}\Big({\mP^2\over\mc\Tx}\Big)
\Big({\Ts\over\mP}\Big)^{5/2}
\Big({\Gam\Ts\,\over\sqrt\aas\,\mP}\Big)^{(3+\p)/(6-2\p)}\ ,
\label{add 18} 
\ee

Let us now again examine some specific examples.

\subsection{Case $\bf \Tx \approx\Ts$.}

The formula (\ref{old 27}) is applicable to the case considered in earlier
work\cite{C}, in which it was supposed that vortons sufficiently stable 
to last until the present epoch were formed as the result of the carrier 
condensation occurring  at or very soon after the time of string formation. 
This example provides the strongest limits on $\Tx$.
Setting $\Ts$ equal to $\Tx$ in
(\ref{old 27}) one obtains

\be
\beta^{5/9}{\Tx\over\mP} \left({\nua\mP\over\aas\mc}\right)^{4/9}
\lta 1\ .
\label{plus 39}
\ee
Substituting the estimates above (supposing, as before, that the efficiency
and drag factors are order unity), 
we obtain

\be
\Tx \lta 10^7\, \GeV\  .
\label{plus 40}
\ee

This result is based on the assumptions that the vortons in question are 
stable enough
to survive until the present day. Thus, this constraint is naturally more 
severe
than its analogue in the previous section. It is to be
remarked that  vortons produced in a phase transition occurring at or near the
limit that has just been derived would give a significant contribution to the
elusive dark matter in the universe. However, if they were produced at the
electroweak scale, i.e. with $\Tx\approx\Ts \approx\TEW$, where $\TEW\approx
10^2\, \GeV$, then they would constitute such a small dark matter fraction,
$\Ov\approx 10^{-9}$, that they would be very difficult to detect.

\subsection{Case $\bf \Tx\simeq\TGUT\approx10^{16}$ GeV.}

For the most commonly considered case, namely that of strings formed during
the GUT transition, the nucleosynthesis limit (\ref{old 26}) is already
sufficient for the exclusion of carrier condensation in the friction damping
regime. To obtain the stronger limit that is applicable if the vortons are
sufficiently stable to survive as a dark matter constituent,
we need to consider the case in which the condensation occurs during the
the regime of gravitational damping. In this case, the relevant dark
matter fraction is given by (\ref{add 18}). Setting $\Tx$ equal to $\TGUT$ in
this formula, and dropping the order of unity coefficients we obtain the 
corresponding limit

\be
{\Ts\over\mP}\lta\Big({\mc\TGUT\over\mP^2}\Big)^{(3-\p)/(9-2\p)} \ .
\label{add 19}
\ee
If the loop production were extremely efficient, $\p\simeq 1$,
this would already give the numerical limit

\be
\Ts\lta 10^{10} \GeV\ .
\label{oldlimit} 
\ee
which is significantly stronger than the more
conservative limit (\ref{old 26}) that pertains if the vortons only survive 
for a few minutes. 

However, contrary to what one might have guessed, the highest conceivable
loop production efficiency is not what maximises ultimate vorton production.
This is because it merely tends to enhance the charge and current loss rate by
production not of protovortons but of doomed loops.
Thus if, instead
of supposing that the loop production efficiency $\p$ is close to a hundred
percent, one makes the plausible supposition that it does not much exceed 
fifty per cent, then one obtains
 
\be
\Ts\lta 10^{9} \GeV\ .
\label{newlimit} 
\ee
This limit is still compatible by a very large margin with one of the most
obvious, albeit rather extreme, possibilities\cite{DPWP} that comes to mind,
namely that in which the strings are formed at the GUT level,
$\Tx\approx\TGUT$, but no current carrier condenses until the electroweak
level, $\Ts\approx\TEW$.

\section{Conclusions}

We have explored the constraints and implications both for particle physics
and cosmology arising from the existence of populations of remnant vortons of
more general types than have previously been considered. Specifically, we have
envisaged the possibility of cosmic string superconductivity by
condensation of the relevant carrier field at energy scales significantly
below that of the string formation. We have seen that there are two
qualitatively very different possibilities. In scenarios for which the carrier
condensation occurs at comparitively high energy, during the friction damping
regime, a substantial majority of the superconducting string loops will
ultimately survive as vortons. Such scenarios are more easily excluded on
observational grounds than the alternative possibility, which is that 
superconductivity
does not set in until a later stage, in which case only a minority of the
loops initially present ultimately become vortons.

We have shown that large classes of particle physics models can provisionally
be ruled out as incompatible by these cosmological considerations,  and in
particular we have shown that models admitting GUT strings must not allow
any string superconductivity giving stable vortons to set in much
above $10^{9}$ GeV. The excluded regions of parameter space are shown in
figure~2.

Our conclusions are, however, dependent on a number of more or less
``conventional" assumptions, whose validity will need to be systematically
scrutinised in future work. Invalidation of these conventional assumptions, 
particularly those concerning the long term stability of the vortons, in
specific theoretical contexts would mean that in such circumstances the
constraints given here might need to be considerably relaxed. On the other
hand, our constraints may be considerably tightened by the use of
more detailed observational data and the ensuing limits on the populations of
various kinds of vortons that can exist today. On the constructive side, we 
have
shown that it is possible for various conceivable symmetry breaking
schemes to give rise to a remnant vorton density sufficient to make up a
significant portion of the dark matter in the universe.

\section{Acknowledgements}
R.B. is supported in part by the US Department of Energy under Grant
\#DE-FG0291ER40688; B.C. is supported by the CNRS; A.D. is supported by PPARC;
and M.T. is supported in part by funds provided by the U.S. Department of
Energy under cooperative research agreement \#DF-FC02-94ER40818.  We have
benefitted from helpful converations with many colleagues including Warren 
Perkins, Patrick Peter, Tsvi Piran, Paul Shellard and Neil Turok.

\begin{center}
\bf Figure Captions
\end{center}
Figure 1: The evolution of relevant physical quantities as a function
of the thermal length scale in the early 
universe.
\vspace{5mm}
\newline
Figure 2:  A summary of our results. This diagram shows how the appropriate
length scales in particle physics theories are constrained. Both the
nucleosynthesis and dark matter bounds are shown.

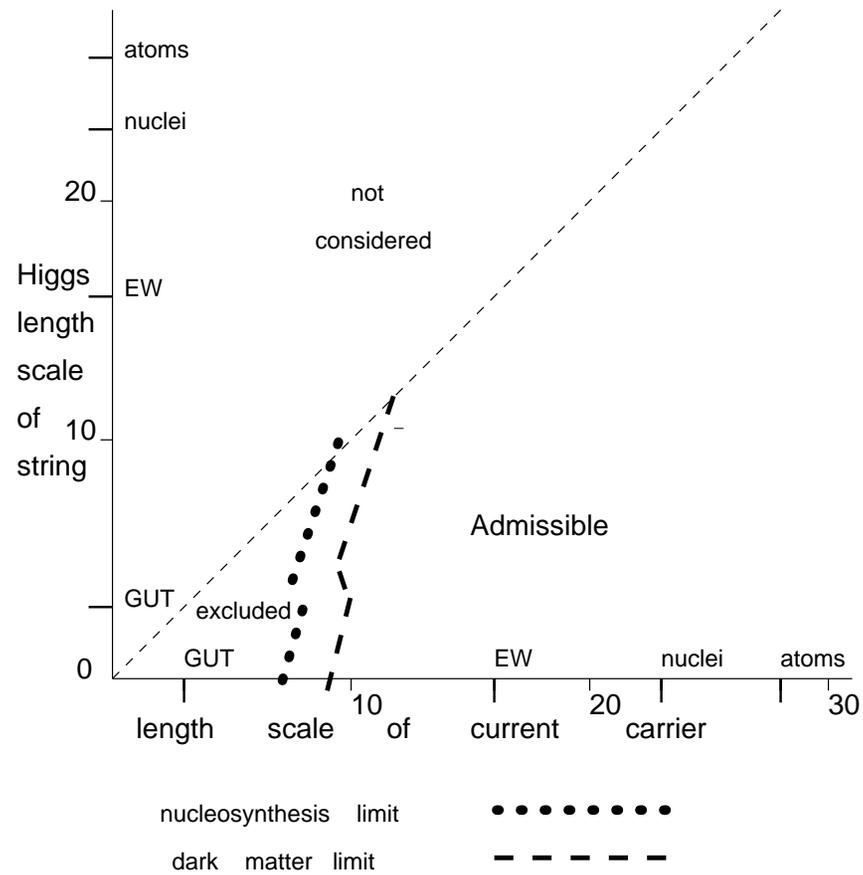
\end{document}